\documentclass[runningheads]{llncs}

\usepackage{algorithm}
\usepackage[noend]{algpseudocode}
\algrenewcommand\algorithmicindent{0.8em}

\usepackage{wrapfig}

\usepackage{etoolbox}
\BeforeBeginEnvironment{wrapfigure}{\setlength{\intextsep}{0pt}}

\usepackage{epstopdf}
\usepackage{framed}

\usepackage{float}

\usepackage{bbding} 

\usepackage{booktabs}
\newcommand{\ra}[1]{\renewcommand{\arraystretch}{#1}}

\usepackage{amssymb}
\newcommand{\pluseq}{\mathrel{+}\mathrel{\mkern-2mu}=}
\newcommand{\minuseq}{\mathrel{-}\mathrel{\mkern-2mu}=}

\usepackage{enumitem}

\usepackage{subcaption}

\usepackage{microtype}

\usepackage{graphicx}
\usepackage{hyperref}
\usepackage{xcolor}

\hypersetup{
    colorlinks,
    linkcolor={red!50!black},
    citecolor={blue!50!black},
    urlcolor={blue!80!black}
}

\usepackage{blindtext} 

\usepackage[subtle]{savetrees}

\usepackage{fixmath}

\begin{document}
\title{S-RASTER: Contraction Clustering for\\Evolving Data Streams\thanks{The final authenticated version is available online at \url{https://doi.org/10.1186/s40537-020-00336-3}.}
}

%
%

\author{Gregor Ulm\inst{1, 2}\Envelope
\orcidID{0000-0001-7848-4883}
\and
Simon Smith\inst{1, 2}
\orcidID{0000-0001-8525-2474}
\and
Adrian Nilsson\inst{1, 2}
\orcidID{0000-0002-8927-845X}
\and
Emil Gustavsson \inst{1, 2}
\orcidID{0000-0002-1290-9989}
\and
Mats Jirstrand \inst{1, 2}
\orcidID{0000-0002-6612-8037}
}
\authorrunning{G. Ulm et al.}

\institute{Fraunhofer-Chalmers Research Centre for Industrial Mathematics,\\ Chalmers Science
Park, 412 88 Gothenburg, Sweden\\
\and Fraunhofer Center for Machine Learning,
\\Chalmers Science
Park, 412 88 Gothenburg, Sweden\\
\email{\{gregor.ulm, simon.smith, adrian.nilsson, \\emil.gustavsson, mats.jirstrand\}@fcc.chalmers.se}\\
\url{http://www.fcc.chalmers.se/}
}

\maketitle              
\begin{abstract}
Contraction Clustering (RASTER) is a single-pass algorithm for density-based clustering of 2D data. It can process arbitrary amounts of data in linear time and in constant memory, quickly identifying approximate clusters. It also exhibits good scalability in the presence of multiple CPU cores. RASTER exhibits very competitive performance compared to standard clustering algorithms, but at the cost of decreased precision. Yet, RASTER is limited to batch processing and unable to identify clusters that only exist temporarily. In contrast, S-RASTER is an adaptation of RASTER to the stream processing paradigm that is able to identify clusters in evolving data streams. This algorithm retains the main benefits of its parent algorithm, i.e.~single-pass linear time cost and constant memory requirements for each discrete time step within a sliding window. The sliding window is efficiently pruned, and clustering is still performed in linear time. Like RASTER, S-RASTER trades off an often negligible amount of precision for speed. Our evaluation shows that competing algorithms are at least 50\% slower. Furthermore, S-RASTER shows good qualitative results, based on standard metrics. It is very well suited to real-world scenarios where clustering does not happen continually but only periodically. 
\end{abstract}

\section{Introduction}
Clustering is a standard method for data analysis and many clustering methods have been proposed~\cite{clusteringSurvey}. Some of the most well-known clustering algorithms are DBSCAN~\cite{ester1996density}, $k$-means clustering~\cite{macqueen1967some}, and CLIQUE~\cite{Agrawal1998}~\cite{Agrawal2005}. Yet, they have in common that they do not perform well with big data, i.e.~data that far exceeds available main memory~\cite{Venkatasubramanian2018}. This was also confirmed by our own experience when we faced the real-world industrial challenge of identifying dense clusters in terabytes of geospatial data. This led us to develop Contraction Clustering (RASTER), a very fast linear-time clustering algorithm for identifying approximate density-based clusters in 2D data, primarily motivated by the fact that existing batch processing algorithms for this purpose exhibited insufficient performance. We previously described RASTER and highlighted its performance for sequential processing of batch data~\cite{rasterLOD}. This was followed by a description of a parallel version of that algorithm~\cite{ulm2019contraction}. A key aspect of RASTER is that it does not exhaustively cluster its input but instead identifies their approximate location in linear time. As it only requires constant space, it is eminently suitable for clustering big data. The variant RASTER$'$ retains its input and still runs in linear time while requiring only a single pass. Of course, it cannot operate in constant memory.

\begin{figure*}
\centering
\begin{subfigure}{.25\textwidth}
  \centering
  \includegraphics[scale=0.40, trim=1.2cm 2.69cm 2.16cm 0.7cm, clip]{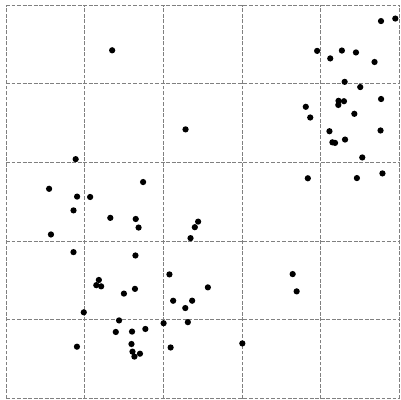}
  \caption{Input data}
  \label{fig:idea1}
\end{subfigure}\hfill
\begin{subfigure}{.25\textwidth}
  \centering
  \includegraphics[scale=0.40, trim=1.2cm 2.69cm 2.16cm 0.7cm, clip]{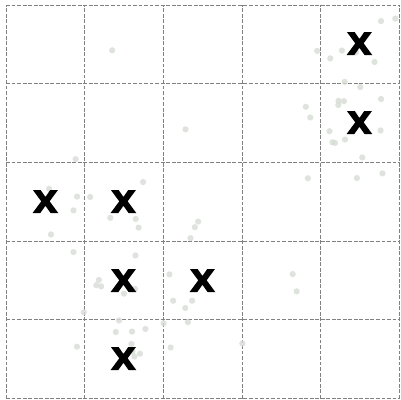}
  \caption{Significant tiles}
  \label{fig:idea2}
\end{subfigure}\hfill
\begin{subfigure}{.25\textwidth}
  \centering
  \includegraphics[scale=0.40, trim=1.2cm 2.69cm 2.16cm 0.7cm, clip]{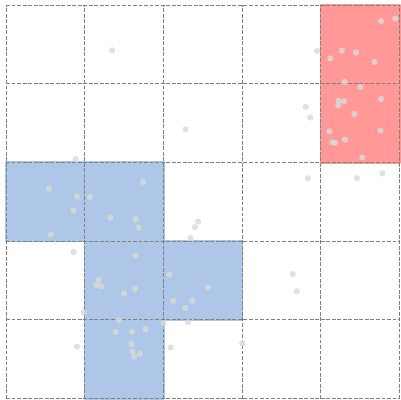}
  \caption{Clustered tiles}
  \label{fig:idea3}
\end{subfigure}\hfill
\begin{subfigure}{.25\textwidth}
  \centering
  \includegraphics[scale=0.40, trim=1.2cm 2.69cm 2.16cm 0.7cm, clip]{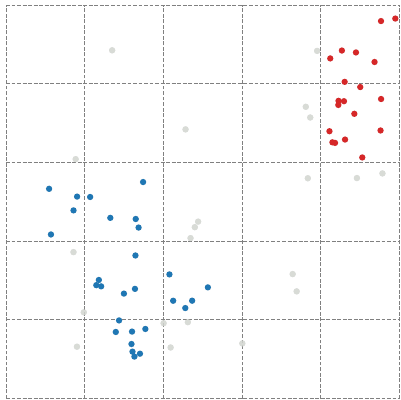}
  \caption{Clustered points}
  \label{fig:idea4}
\end{subfigure}
\caption{High-level visualization of RASTER (best viewed in color), using a simplified example on a small 5x5 grid. Please also refer to Sect.~\ref{bg-raster} and Table~\ref{table:symbols}. The precision of the input is reduced, with leads to an implied grid. This grid is shown to aid the reader but it is not explicitly constructed by the algorithm. The original input is shown in Fig.~\ref{fig:idea1}, followed by projection to tiles in Fig.~\ref{fig:idea2} where only significant tiles that contain at least $\tau = 4$ values are retained. Tiles that contain less than $\tau$ values are subsequently ignored as they are treated as noise. The result is a set of significant tiles. The parameter $\mu$ specifies how many significant tiles a cluster has to contain as a minimum. In this case, given a minimum cluster size of $\mu = 2$ and a maximum distance $\delta = 1$, i.e.~significant tiles need to be adjacent, two clusters emerge (cf.~Fig.~\ref{fig:idea3}), which corresponds to RASTER. Clusters as collections of points are shown in Fig.~\ref{fig:idea4}, which corresponds to the variant RASTER$'$.}
\label{fig:idea}
\end{figure*}

A common motivation for stream processing is that data does not fit into working memory and therefore cannot be retained. This is not a concern for RASTER as it can process an arbitrary amount of data in limited working memory. One could therefore divide a stream of data into discrete batches and consecutively cluster them. Yet, this approach does not address the problem that, in a given stream of data, any density-based cluster may only exist temporarily. In order to solve this problem, this paper presents Contraction Clustering for Evolving Data Streams (S-RASTER). This algorithm has been designed for identifying density-based clusters in infinite data streams within a sliding window. S-RASTER is not a replacement of RASTER, but a complement, enabling this pair of algorithms to efficiently cluster data, regardless of whether it is available as a batch or a stream.

Given that there is already a number of established algorithms for detecting clusters in data streams, the work on S-RASTER may need to be further motivated. The original motivation is related to the batch processing clustering algorithm RASTER, which is faster than competing algorithms and also requires less memory. This comes at the cost of reduced precision, however. For the use case we have been working on, i.e.~very large data sets at the terabyte level, even modest improvements in speed or memory requirements compared to the status quo were a worthwhile pursuit. Indeed, RASTER satisfied our use case in industry as it is faster than competing algorithms and also requires less memory. Yet, we wanted to improve on it as it did not help us to process data streams. It also is not able to detect clusters that only exist temporarily. Furthermore, we also wanted to explore if there is a faster way to process the data we needed to process, given our constraints (cf.~Sect.~\ref{bg-problem}). These reasons motivated our work on S-RASTER. The goal was to achieve faster data processing than existing methods allow, using only a single pass, but with an acceptable loss of precision. As the results in this paper show, S-RASTER is, in a standard benchmark, indeed faster than competing algorithms for clustering data streams.

\begin{figure}[t]
\centering
\includegraphics[scale=0.48, trim= 5.05cm 0.0cm 1.2cm 0.0cm, clip]{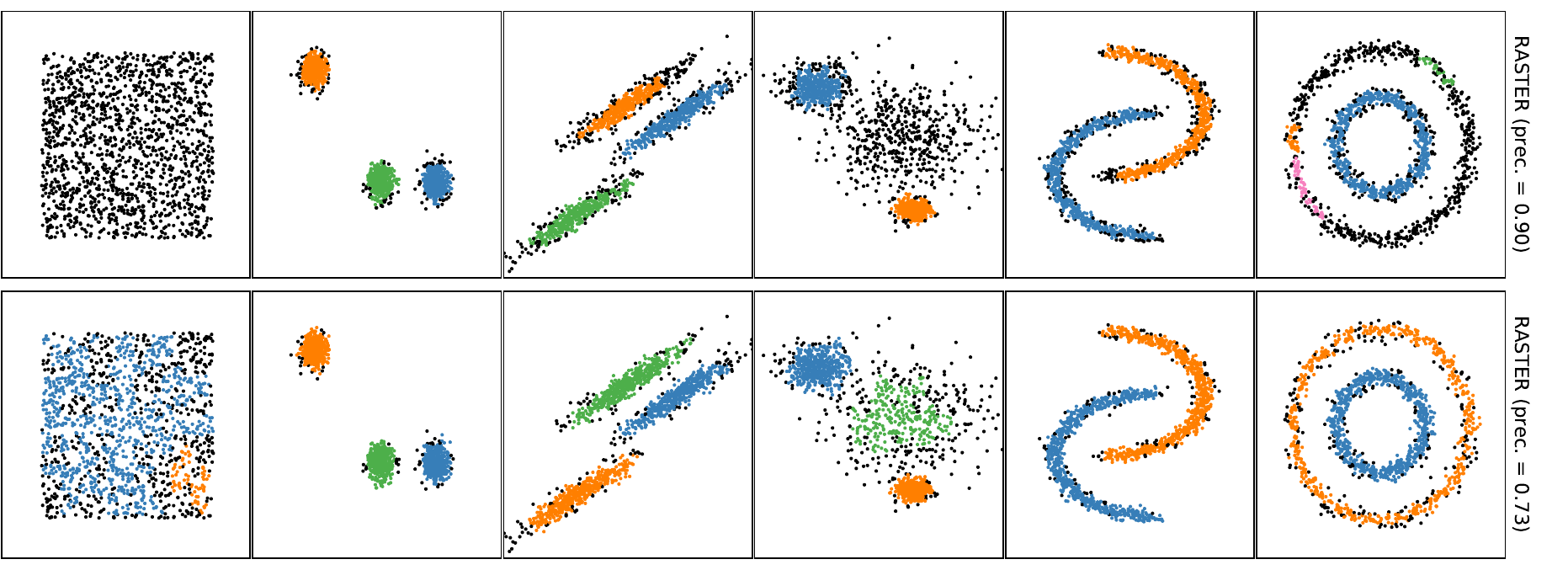}
\caption{The precision parameter $\xi$ greatly influences clustering results of RASTER (best viewed in color). This illustration is based on retaining all data points (cf.~Fig.~\ref{fig:idea4}). With a precision of $\xi = 0.90$ (top), all but the rightmost data set are clustered satisfactorily. Reducing the precision to $\xi = 0.73$ (bottom) improves the results of that data set. It is a matter of debate which value of $\xi$ led to a better result with the data set in the middle as a good case could be made for either result, depending on whether the goal of the user is to identify dense or sparse clusters. The data sets were taken from a collection of standard data sets for the evaluation of general-purpose clustering algorithms that are part of the machine learning library \texttt{scikit-learn}~\cite{pedregosa2011scikit}. A more extensive discussion of these results is provided in a previous paper on RASTER~\cite{ulm2019contraction}.}
\label{fig:rvsr}
\end{figure}

In the remainder of this paper, we provide relevant background in Sect.~\ref{sec-background}, which contains a brief recapitulation of RASTER and the motivating use case for S-RASTER, i.e.~identifying evolving hubs in streams of GPS data. In Sect.~\ref{sec-sraster} we provide a detailed description of S-RASTER, including a complete specification in pseudocode. This is followed by a theoretical evaluation of S-RASTER in Sect.~\ref{eval-theory} and a description of our experiments in Sect.~\ref{sec-exp}. The results of our experiments as well as a discussion of them are presented in Sect.~\ref{sec-eval}. Some related work is part of the evaluation section as we compare S-RASTER to competing algorithms. However, further related work is highlighted in Sec.~\ref{sec-related} and future work in Sec.~\ref{sec-future}. We finish with a conclusion in Sect.~\ref{sec-conclusion}.

\section{Background}
\label{sec-background}
In this section, we give a brief presentation of the sequential RASTER algorithm in Subsect.~\ref{bg-raster}. This is followed by a description of the motivating problem behind S-RASTER, i.e.~the identification of so-called hubs within a sliding window, in Subsect.~\ref{bg-problem}.

\subsection{RASTER}
\label{bg-raster}

In this subsection, we provide a brief description of RASTER~\cite{rasterLOD}~\cite{ulm2019contraction}. This algorithm approximately identifies density-based clusters very quickly (cf.~Alg.~\ref{alg:raster}). The main idea is to project data points to tiles and keep track of the number of points that are projected to each tile. Only tiles to which more than a predefined threshold number~$\tau$ of data points have been projected are retained. These are referred to as significant tiles $\sigma$, which are subsequently clustered by exhaustive lookup of neighboring tiles in a depth-first manner. Clustering continues for as long as there are significant tiles left. To do so, the algorithm selects an arbitrary tile as the seed of a new cluster. This cluster is grown iteratively by looking up all neighboring tiles within a given Manhattan or Chebyshev distance $\delta$. This takes only $\mathcal{O}(1)$ as the location of all potential neighbors is known due to their location in the grid. Only clusters that contain more than a predefined number $\mu$ of significant tiles are retained. While RASTER was developed to identify dense clusters, it can identify clusters that have irregular shapes as well. Refer to Fig.~\ref{fig:rvsr} for an illustration.

The projection operation consists of reducing the precision of the input by scaling a floating-point number to an integer. For instance, take an arbitrary GPS coordinate $(34.59204302, 106.36527351)$, which is already truncated compared to the full representation with double-precision floating-point numbers. GPS data is inherently imprecise, yet stored in floating-point format with the maximum precision, which is potentially misleading, considering that consumer-grade GPS is only accurate to within about five to ten meters in good conditions, primarily open landscapes~\cite{diggelenGPS}~\cite{WingGPS}. In contrast, in urban landscapes GPS accuracy tends to be much poorer. A study based on London, for instance, found a mean error of raw GPS measurements of vehicular movements of 53.7m~\cite{TaylorGPS}. Consequently, GPS data suggests a level of precision they do not possess. Thus, by dropping a few digit values, we do not lose much, if any, information. Furthermore, vehicle GPS data is sent by vehicles that may, in the case of trucks with an attached trailer, be more than 25 meters long. To cluster such coordinates, we can truncate even more digit points. For instance, if four place values after the decimal point are enough, which corresponds to a resolution of 11.1 meters in the case of GPS, we transform the aforementioned sample data point to $(34.5920, 106.3652)$. However, to avoid issues pertaining to working with floating-point numbers, the input is instead scaled to $(345920, 1063652)$. Only before generating the final output, all significant tiles of the resulting clusters are scaled back to floating-point numbers, i.e.~the closest floating-point representation of $(34.592, 106.3652)$.

\algrenewcommand\algorithmicrequire{\textbf{input:}}
\algrenewcommand\algorithmicensure{\textbf{output:}}
\begin{wrapfigure}{R}{0.60\textwidth}
\begin{minipage}{0.60\textwidth}
\begin{algorithm}[H]
\begin{algorithmic}[1]
\Require data \emph{points}, precision $\xi$, threshold $\tau$, distance $\delta$, minimum cluster size $\mu$
\Ensure set of clusters \emph{clusters}
\State $\mathit{acc} := \varnothing $ \Comment $\{(x_\pi, y_\pi): \mathit{count}\}$ 
\State $\mathit{clusters} := \varnothing$ \Comment set of sets

\For{$(x, y)$ in \emph{points}} 
\State $(x_\pi, y_\pi) := \mathit{project}(x, y, \xi)$ \Comment $\mathcal{O}(1)$
\If{$(x_\pi, y_\pi)$ $\not\in$ keys of \emph{acc}}{}
\State $\mathit{acc}[(x_\pi, y_\pi)] := 1$
\Else
\State $\mathit{acc}[(x_\pi, y_\pi)] \pluseq 1$
\EndIf
\EndFor

\For{$(x_\pi, y_\pi)$ in $\mathit{acc}$}
\If{$\mathit{acc}[(x_\pi, y_\pi)] < \tau$}
\State remove $\mathit{acc}[(x_\pi, y_\pi)]$
\EndIf
\EndFor

\State $\sigma := $ keys of $\mathit{acc}$ \Comment significant tiles

\While{$\sigma \ne\varnothing$} \Comment $\mathcal{O}(n)$ for lls. 12--24
\State $t$ := $\sigma\mathit{.pop()}$
\State \emph{cluster} := $\varnothing$ \Comment set
\State \emph{visit} := $\{t\}$

\While{\emph{visit} $\ne\varnothing$}
  \State $u := $ $\mathit{visit.pop()}$
  \State $\mathit{ns} := \mathit{neighbors}(u, \delta)$  \Comment $\mathcal{O}(1)$
  \State $\mathit{cluster} := \mathit{cluster} \cup \{u\}$
  \State $\sigma := \sigma \setminus \mathit{ns}$  \Comment cf.~ln.~13
  \State $\mathit{visit} := \mathit{visit} \cup \mathit{ns}$
\EndWhile

\If{size of $\mathit{cluster} \geq \mu$}
\State add $\mathit{cluster}$ to $\mathit{clusters}$
\EndIf
\EndWhile

\caption{\textsc{RASTER}}
\label{alg:raster}
\end{algorithmic}
\end{algorithm}
\end{minipage}
\end{wrapfigure}

Two notes regarding the reduction of precision are in order. First, this procedure is not limited to merely dropping digit values. As those values are only intermediary representations that are used for clustering, any non-zero real number can be used as the scaling factor. Second, the magnitude of the scaling factor depends on two aspects, precision of the provided data and size of the objects we want to identify clusters of. For the former, it should be immediately obvious that data that suggests a greater precision than it actually possesses, like GPS data mentioned above, can be preprocessed accordingly without any loss of information. The latter depends on the domain and the trade-off the user is willing to make as both clustering speed and memory requirements directly depend on the chosen precision.

As clustering belongs to the field of unsupervised learning, there are potentially multiple satisfactory clusterings possible with any given dataset. With RASTER, the user can influence clustering results by adjusting four parameters: precision $\xi$, threshold for a significant tile $\tau$, maximum distance of significant tiles in a cluster $\delta$, and threshold for the cluster size $\mu$. The precision parameter $\xi$ directly influences the granularity of the implied grid (cf.~Fig.~\ref{fig:idea}). A lower precision value for $\xi$ leads to a coarser grid, and vice versa. With the value $\tau$, it is possible to directly influence the number of significant tiles that are detected in any given data set. The higher this value, the fewer significant tiles will be identified. While the most intuitive value for the distance parameter $\delta$ is 1, meaning that all significant tiles that are combined when constructing a cluster need to be direct neighbors, it is possible to also take the case of sparser clusters into account. For instance, if the user detects two dense clusters that are only one tile apart, it may make sense to set $\delta = 2$ to combine them. Of course, this depends on the application domain. Lastly, the parameter $\mu$ determines when a collection of significant tiles is considered a cluster. The higher this value, the fewer clusters will be detected. 

RASTER is a single-pass linear time algorithm. However, in a big data context, its biggest benefit is that it only requires constant memory, assuming a finite range of inputs. This is the case with GPS data. It is therefore possible to process an arbitrary amount of data on a resource-constrained workstation with this algorithm. We have also shown that it can be effectively parallelized~\cite{ulm2019contraction}. A variation of this algorithm that retains its inputs is referred to as RASTER$'$. It is less suited for big data applications. However, it is effective for general-purpose density-based clustering and very competitive compared to standard clustering methods; cf.~Appendix A in~\cite{ulm2019contraction}.

\begin{figure}
\centering
\includegraphics[
  scale=.65,
  trim=1.3cm 1.3cm 1.3cm 2.2cm, clip]{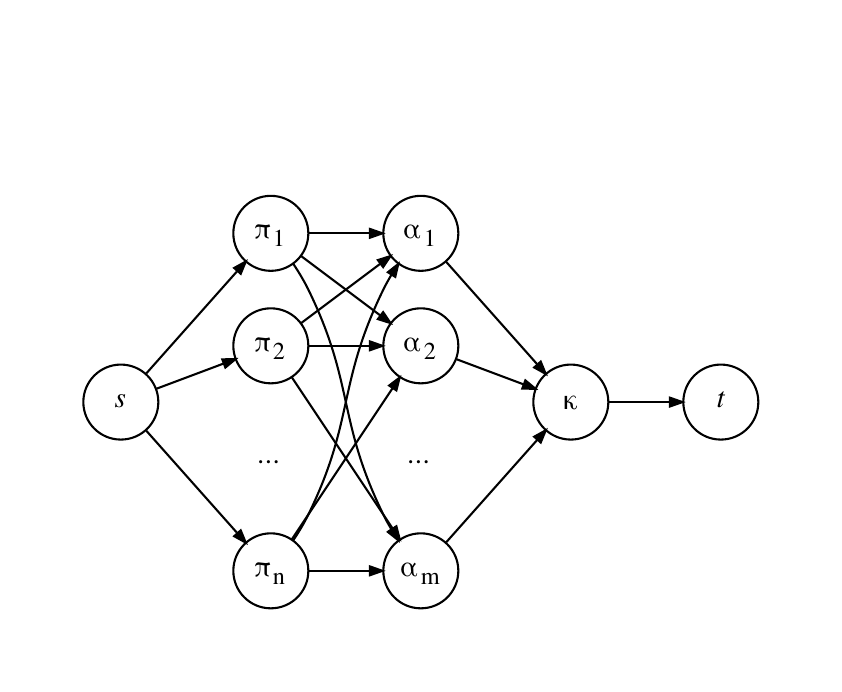}
\caption{
Flow graph of S-RASTER, a modification of RASTER (cf.~Fig.~\ref{fig:idea}) for evolving data streams. Refer to Sect.~\ref{sec-idea} for a detailed description. The input source node $s$ distributes values arbitrarily to projection nodes $\mathbold{\pi}$, which reduce the precision of the input. In turn, they send projected values to accumulation nodes $\mathbold{\alpha}$. These nodes keep a count of points for each tile and determine significant tiles for the chosen sliding window. Should a tile become significant or a once significant tile no longer be significant, a corresponding update is sent to the clustering node $\kappa$. Node $\kappa$ periodically performs clustering of significant tiles, which is a very fast operation.}
\label{fig:graph}
\end{figure}

\subsection{Identifying Evolving Hubs}
\label{bg-problem}
RASTER was designed for finite batches of GPS traces of commercial vehicles. The goal was to identify hubs, i.e. locations where many vehicles come to a halt, for instance vans at delivery points or buses at bus stops. After identifying all hubs in a data set, it is possible to construct vehicular networks. However, what if the data does not represent a static reality? It is a common observation that the location of hubs changes over time. A bus stop may get moved or abolished, for instance. This motivates modifying RASTER so that it is able to detect hubs over time and maintaining hubs within a sliding window $W$. The length of $W$ depends on the actual use case. With GPS traces of infrequent but important deliveries, many months may be necessary. Yet, with daily deliveries, a few days would suffice to detect new hubs as well as discard old ones.

\begin{table*}[t]
\caption{Overview of symbols used in the description of RASTER and S-RASTER. The parameters $\xi$, $\tau$, $\delta$, and $\mu$ are used for both RASTER and S-RASTER and directly influence clustering results. In contrast, the symbols $\pi$, $\alpha$, and $\kappa$ are only used for the description of S-RASTER.} 
\vspace{0.3em}
\centering\ra{1.0}\begin{tabular}{@{}ccl@{}}
\toprule
Symbol & \phantom{ab} & Meaning
 \\
\toprule
$\xi$ & & Precision for projection operation\\
$\tau$ & & Threshold number of points to determine if a tile is significant\\
$\delta$ & & Distance metric for cluster definition\\
$\mu$ & & Minimum cluster size in terms of the number of significant tiles\\
$\pi$ & & Projection operator\\
$\alpha$ & & Accumulation operator\\
$\kappa$ & & Clustering operator\\
\bottomrule\end{tabular}
\label{table:symbols}
\end{table*}

\section{S-RASTER}
\label{sec-sraster}
This section starts with a concise general description of S-RASTER in Sect.~\ref{sec-idea}, followed by a detailed specification in Sect.~\ref{sec-alg}. Afterwards, we highlight some implementation details in Sect.~\ref{sec-implementation} and outline, in Sect.~\ref{sec-prime}, how S-RASTER has to be modified to retain its inputs, which makes this algorithm applicable to different use cases.

\subsection{Idea}
\label{sec-idea}
Before we present a more technical description of S-RASTER, we would like to start with a more intuitive description of this algorithm, using GPS coordinates for the purpose of illustration. For the sake of simplicity, we also ignore parallel computations for the time being. Imagine five sequential nodes: source ($s$), projection ($\pi$), accumulation ($\alpha$), clustering ($\kappa$), and sink ($t$). The general idea is that S-RASTER processes input data using those five nodes. Input is provided by the source $s$ and, without modification, forwarded to node $\pi$, which performs projections, e.g.~it reduces the precision of the provided input by scaling it. A simple example consists of dropping place values of GPS coordinates~(cf.~Sect.~\ref{bg-raster}). The projected values are sent from node $\pi$ to node $\alpha$, which keeps track of the number of points that were projected to each tile in the input space for the duration of the chosen window. Once the status of a tile changes, i.e.~it becomes significant or was once significant but no longer is, an update is sent to node $\kappa$. These steps happen continually whenever there is new data to process. In contrast, node $\kappa$ performs clustering based on significant tiles in a fixed interval, which is followed by sending clusters as sets of significant tiles to the sink node $t$.

More formally, S-RASTER (cf.~Fig.~\ref{fig:graph}) performs, for an indefinite amount of time, projection and accumulation continually, and clustering periodically. Projection nodes $\mathbold{\pi}$ receive their input from the source node $s$. Each incoming pair $(x, y)$, where $x, y \in \mathbb{R}$, is surjected to $(x_\pi, y_\pi)$, where $x_\pi, y_\pi \in \mathbb{Z}$. Together with a period indicator $\Delta_z$, these values are sent to accumulation nodes $\mathbold{\alpha}$. The identifier $\Delta_z \in \mathbb{N}_0$ designates a period with a fixed size, e.g.~one day, and is non-strictly increasing. Each $\alpha$-node maintains a sliding window $W$ of length $c$, which is constructed from $c$ multisets $W_{\Delta_z}$. Each such multiset $W_{\Delta_z}$ keeps running totals of how many times the input was surjected to any given tuple $(x_\pi, y_\pi)$ in the chosen period. The sliding window starting at period $\Delta_i$ is defined as $W_{\Delta_i}^{\Delta_{i+c}} = \bigcup_{z = i}^{i + c} W_{\Delta_z}$.\footnote{The notation $W_{\Delta_i}^{\Delta_{i+c}}$ expresses a window of length $c$, which implies that the upper delimiter is excluded. This is done to simplify the notation as we otherwise would have to specify the upper limit as $i + c - 1$.} It contains the set of significant tiles $\sigma = \{(x_\pi, y_\pi)^d \in W_{\Delta_i}^{\Delta_{i+c}} \mid d \ge \tau \}$, where $d$ indicates the multiplicity, i.e.~the number of appearances in the multiset, and $\tau$ the threshold for a significant tile. If a tile becomes significant, its corresponding value $(x_\pi, y_\pi)$ is forwarded to the clustering node $\kappa$. Whenever $W$ advances from $W_{\Delta_i}^{\Delta_{i+c}}$ to $W_{\Delta_{i+1}}^{\Delta_{i+c+1}}$,  the oldest entry~$W_{\Delta_i}$ is removed from $W$. Furthermore, all affected running totals are adjusted, which may lead to some significant tiles no longer being significant. If so, node $\kappa$ receives corresponding updates to likewise remove those entries. Node $\kappa$ keeps track of all significant tiles, which it clusters whenever $W$ advances (cf.~Alg.~\ref{alg:raster}, lls.~12--24). The preliminary set of clusters is $\mathit{ks}$. The final set of clusters is defined as $\{k \in \mathit{ks} \mid k > \mu \}$, where $\mu$ is the minimum cluster size. Each $(x_\pi, y_\pi)$ in each $k$ is finally projected to $(x_\pi', y_\pi')$, where $x_\pi', y_\pi' \in \mathbb{R}$. Together with cluster and period IDs, these values are sent to the sink node $t$.

\subsection{Detailed Description}
\label{sec-alg}
S-RASTER processes the data stream coming from source $s$ in Fig.~\ref{fig:graph} and outputs clusters to sink $t$. There are three different kinds of nodes: projection nodes $\mathbold{\pi}$ project points to tiles, accumulation nodes $\mathbold{\alpha}$ determine significant tiles within each sliding window $W_{\Delta_i}^{\Delta_{i+c}}$, and one clustering node $\kappa$ outputs, for each $W_{\Delta_i}^{\Delta{i+c}}$, all identified clusters. Below, we describe the three nodes in detail.

\begin{wrapfigure}{R}{0.55\textwidth}
\begin{minipage}{0.55\textwidth}
\begin{algorithm}[H]
\caption{Projection node $\pi$}
\label{alg:project}
\begin{algorithmic}[1]
\Require stream of tuples $(x, y, \Delta_z)$ where $x,y$ are coordinates and $\Delta_z$ is the current period; precision $\xi$ 
\Ensure stream of tuples $(x_\pi, y_\pi , \Delta_z)$
\State $x_\pi := 10^{\xi}x$
\State $y_\pi := 10^{\xi}y$
\State send $(x_\pi, y_\pi, \Delta_z)$ to node $\alpha$
\end{algorithmic}
\end{algorithm}
\end{minipage}
\end{wrapfigure}

\paragraph{Projection.}
Nodes labeled with $\mathbold{\pi}$ project incoming values to tiles with the specified precision (cf.~Alg.~\ref{alg:project}). In general, the input $(x, y, \Delta_z)$ is transformed into $(x_\pi, y_\pi, \Delta_z)$. Projection of any value $v$ to $v_{\pi}$, using precision $\xi$, is defined as $v_{\pi} = 10^{\xi}v$, where $v, \xi \in \mathbb{R}, v_{\pi} \in \mathbb{N}$. This entails that $v_{\pi}$ is forcibly typecast to an integer. Thus, the fractional part of the scaled input is removed as this is the information we do not want to retain. The period $\Delta_z$ is a non-strictly increasing integer, for instance uniquely identifying each day. Projection is a stateless operation that can be executed in parallel. It is irrelevant which node $\pi$ performs projection on which input value as they are interchangeable. However, the input of the subsequent accumulator nodes $\mathbold{\alpha}$ needs to be grouped by values, for instance by assigning values within a certain segment of the longitude range of the input values. Yet, this is not strictly necessary as long as it is ensured that every unique surjected value $(x_\pi, y_\pi)$ is sent to the same $\alpha$-node.

\paragraph{Accumulation.}
The accumulator nodes $\mathbold{\alpha}$ keep track of the number of counts per projected tile in the sliding window $W_{\Delta_i}^{\Delta_{i+c}}$ (cf.~Alg.~\ref{alg:count}). The input consists of a stream of tuples $(x_\pi, y_\pi, \Delta_z)$ as well as the size of the sliding window $c$ and the threshold value $\tau$. A global variable $\Delta_j$ keeps track of the current period. Each $\alpha$-node maintains two persistent data structures. For reasons of efficiency, the hashmap \emph{totals} records, for each tile, how many points were projected to it in $W_{\Delta_i}^{\Delta_{i+c}}$. Tiles become significant once their associated count reaches $\tau$. In addition, the hashmap \emph{window} records the counts per tile for each period $W_{\Delta_z}$ in $W_{\Delta_i}^{\Delta_{i+c}}$. Given that input $\Delta_z$ is non-strictly increasing, there are two cases to consider:

\begin{enumerate}[label=(\roman*)]
\item $\Delta_z = \Delta_j$, i.e. the period is unchanged. In this case, the count of tile $(x_\pi, y_\pi)$ in \emph{totals} as well as its count corresponding to the key $\Delta_z$ in \emph{window} are incremented by 1. If the total count for $(x_\pi, y_\pi)$ has just reached $\tau$, an update is sent to the $\kappa$-node, containing $(x_\pi, y_\pi)$ and the flag 1.
\item $\Delta_z > \Delta_j$, i.e.~the current input belongs to a later period. Now the sliding window needs to be advanced, which means that the oldest entry gets pruned. But first, an update is sent to the $\kappa$-node with $\Delta_j$ and the flag 0. Afterwards, the entries in the hashmap \emph{window} have to be adjusted. Entry $\Delta_{z - c}$ is removed and for each coordinate pair and its associated counts, the corresponding entry in the hashmap \emph{totals} gets adjusted downward as the totals should now no longer include these values. In case the associated value of a coordinate pair in \emph{totals} drops below $\tau$, an update is sent to~$\kappa$, consisting of $(x_\pi, y_\pi)$ and the flag -1. Should a value in \emph{totals} reach 0, the corresponding key-value pair is removed. Afterwards, the steps outlined in the previous case are performed.
\end{enumerate}

Regarding significant tiles, only status changes are communicated from $\mathbold{\alpha}$ nodes to $\kappa$, which is much more efficient than sending a stream with constant status updates for each tile.

\begin{wrapfigure}{R}{0.62\textwidth}
\begin{minipage}{0.62\textwidth}
\begin{algorithm}[H]
\caption{Accumulation node $\alpha$}
\label{alg:count}
\begin{algorithmic}[1]
\Require stream $s$ of tuples $(x_\pi, y_\pi, \Delta_z)$, sliding window size $c$, threshold $\tau$

\Ensure stream of tuples  $(\mathit{flag}, \mathit{val})$ where \emph{flag} $\in \{-1,0,1\}$ and \emph{val} $\in \{(x_\pi, y_\pi), \Delta_j\}$ 
\State $\mathit{totals} := \varnothing$
  \Comment{ $\{(x_\pi, y_\pi): \mathit{count}}\}$
\State $\mathit{window} := \varnothing$
  \Comment{ $\{\Delta_j : \{(x_\pi, y_\pi): \mathit{count}\}\}$}
\State $\Delta_j := -1$ \Comment Period count 
\For{$(x_\pi, y_\pi, \Delta_z)$ in $s$} \Comment $\Delta_z - \Delta_j \in \{0,1\}$
\If{$\Delta_z > \Delta_j $}  \Comment New period
  \State send $(0, \Delta_j)$ to node $\kappa$ \Comment Re-cluster
  \State $\mathit{\Delta_j} := \Delta_z$
  \State $\mathit{key}$ := $\Delta_j - c$ \Comment Oldest Entry
\If{$\mathit{key} \in$ keys of $\mathit{window}$} \Comment Prune
  \State $\mathit{vals}$ := $\mathit{window}[\mathit{key}]$
  \For{$(a, b)$ in keys of $\mathit{vals}$}
    \State{$\mathit{old} := \mathit{totals}[(a,b)]$}
    \State{$\mathit{totals}[(a,b)] \minuseq \mathit{vals}[(a,b)]$}
    \State{$\mathit{new} := \mathit{totals}[(a,b)]$}
      
    \If{$\mathit{old} \ge \tau$ and $\mathit{new} < \tau$}
    \State{send $({-1}, (a, b))$ to node $\kappa$ \Comment Remove}
    \EndIf
    \If{$\mathit{new} = 0$}
    \State{remove entry $(a,b)$ from $\mathit{totals}$}
    \EndIf 
       
    \EndFor
  \State{remove entry $\mathit{key}$ from $\mathit{window}$}
\EndIf

\EndIf

\If{$(x_\pi, y_\pi)$ $\notin$ keys of $\mathit{totals}$}
  \State $\mathit{totals}[(x_\pi, y_\pi)] := 1$
  \State $\mathit{window}[\Delta_z][(x_\pi, y_\pi)] := 1$ 
\Else
  \State $\mathit{totals}[(x_\pi, y_\pi)] \pluseq 1$
  \If{$(x_\pi, y_\pi)$ $\notin$ keys of $\mathit{window}[\Delta_z]$}
  \State $\mathit{window}[\Delta_z][(x_\pi, y_\pi)] := 1$
  \Else
  \State $\mathit{window}[\Delta_z][(x_\pi, y_\pi)] \pluseq 1$
  \EndIf  
\EndIf

\If{$\mathit{totals}[(x_\pi, y_\pi)] = \tau$}
  \State send $(1, (x_\pi, y_\pi))$ to node $\kappa$ \Comment Add
\EndIf

\EndFor 
\end{algorithmic}
\end{algorithm}
\end{minipage}
\end{wrapfigure}

\paragraph{Clustering.}
The clustering node $\kappa$ (cf.~Alg.~\ref{alg:cluster}) takes as input a precision value $\xi$, which is identical to the one that was used in the $\alpha$-node, the minimum cluster size $\mu$, and a stream consisting of tuples of a flag $\in \{-1,0,1\}$  as well as a value $\mathit{val}$ $\in \{(x_\pi, y_\pi),  \Delta_j\}$. This node keeps track of the significant tiles~$\sigma$ of the current sliding window, based on updates received from all $\mathbold{\alpha}$ nodes. If $\mathit{flag} = 1$, the associated coordinate pair is added to $\sigma$. On the other hand, if $\mathit{flag} = -1$, tile $(x_\pi, y_\pi)$ is removed from $\sigma$. Thus, $\sigma$ is synchronized with the information stored in all $\mathbold{\alpha}$ nodes. Lastly, if $\mathit{flag} = 0$, the associated value of the input tuple represents a period identifier $\Delta_j$. This is interpreted as the beginning of this period and, conversely, the end of period $\Delta_{j-1}$. Now $\kappa$ clusters the set of significant tiles (cf.~Alg.~\ref{alg:raster}, lls.~12 -- 24), taking $\mu$ into account, and produces an output stream that represents the clusters found within the current sliding window. In this stream, each projected coordinate $(x_\pi, y_\pi)$ is assigned period and cluster identifiers. The coordinate pairs $(x_\pi, y_\pi)$ are re-scaled to floating point numbers $(x_\pi', y_\pi')$ by reversing the operation performed in node $\pi$ earlier. The output thus consists of a stream of tuples of the format $(\Delta_j, \mathit{cluster\_id}, x_\pi', y_\pi')$.

\begin{wrapfigure}{R}{0.70\textwidth}
\begin{minipage}{0.70\textwidth}

\begin{algorithm}[H]
\caption{Clustering node $\kappa$}
\label{alg:cluster}
\begin{algorithmic}[1]
\Require stream $s$ of tuples $(\mathit{flag}, \mathit{val})$ where \emph{flag} $\in \{-1,0,1\}$ and \emph{val} $\in \{(x_\pi, y_\pi),  \Delta_j\}$, precision $\xi$, size $\mu$
\Ensure stream of tuples $(\Delta_j, \mathit{cluster\_id}, x_\pi', y_\pi')$
\State $\sigma := \varnothing$ \Comment Significant tiles
\For{$(\mathit{flag}, \mathit{val})$ in $s$}
  \If{$\mathit{flag} = 1$}
    \State $(x_\pi, y_\pi) := \mathit{val}$
    \State $\sigma := \sigma \cup \{(x_\pi, y_\pi)\}$
  \EndIf
  \If{$\mathit{flag} = -1$}
    \State $(x_\pi, y_\pi) := \mathit{val}$
    \State $\sigma := \sigma \setminus \{(x_\pi, y_\pi)\}$
  \EndIf
  \If{$\mathit{flag} = 0$} \Comment Next period
    \State $\Delta_j := \mathit{val}$
    \State $\mathit{clusters}$ := cluster($\sigma, \mu$) \Comment cf.~Alg.~\ref{alg:raster}, lls.~12 -- 24 
    
    \State $\mathit{id} := 0$ \Comment{Cluster ID}
    \For{cluster in clusters}
      \For{$(x_\pi, y_\pi)$ in cluster}
        \State $(x_\pi', y_\pi') := \mathit{rescale}(x_\pi, y_\pi, \xi)$ \Comment Int to Float
        \State send $(\Delta_j, \mathit{id}, x_\pi', y_\pi')$ 
      \EndFor
      \State $\mathit{id}\pluseq 1$
    \EndFor
  \EndIf
\EndFor
\end{algorithmic}
\end{algorithm}
\end{minipage}
\end{wrapfigure}

\subsection{Implementation Details}
\label{sec-implementation}
 The previous description is idealized. Yet, our software solution has to take the vagaries of real-world data into account. Below, we therefore highlight two relevant practical aspects that the formal definition of our algorithm does not capture.
 
\paragraph{Out-of-Order Processing.}
Tuples are assigned a timestamp at the source, which is projected to a period identifier $\Delta_z$. Assuming that the input stream is in-order, parallelizing the $\alpha$-operator could nonetheless lead to out-of-order input of some tuples at the $\kappa$ node, i.e.~the latter could receive a notification about the start of period $\Delta_j$, cluster all significant tiles as they were recorded up to the seeming end of $\Delta_{j-1}$, but receive further tuples pertaining to it from other $\alpha$-nodes afterwards. One solution is to simply ignore these values as their number should be minuscule. Commonly used periods, e.g.~one day, are quite large and the expected inconsistencies are confined to their beginning and end. Thus, it may make sense to set the start of a period to a time where very little data is generated, e.g.~3:00~a.m.~for a 24-hour period when processing data of commercial vehicles. Depending on actual use cases, other engineering solutions may be preferable. One promising approach would be to not immediately send a notification to initiate clustering to node $\kappa$ when the first element of a new period $\Delta_j$ is encountered by any $\alpha$ node. Instead, one could buffer a certain number of incoming elements that are tagged with $\Delta_j$ for a sufficiently large amount of time, call it a grace period. During that time, elements tagged with $\Delta_j$ are not processed yet. Instead, only elements that are tagged with $\Delta_{j-1}$ are. Once the grace period is up, a signal is sent to $\kappa$ to initiate clustering and the buffered elements in each $\alpha$ node are processed.

\paragraph{Interpolating Periods.} The algorithm as it is described does not take into account that there could be periods without new data. For instance, after processing the last data point in period $\Delta_x$, the next data point may belong to period $\Delta_{x+2}$. When pruning the sliding window, it is therefore not sufficient to only remove data for key $\Delta_{x+2-c}$, where $c$ is the size of the sliding window, as this would lead to data for $\Delta_{x+1-c}$ remaining in perpetuity. Thus, the sliding window also has to advance when there is no data for an entire period. If a gap greater than one period between the current and the last encountered period is detected, the algorithm has to advance the sliding window as many times as needed, one period at a time. After each period, the $\kappa$ node then clusters the significant tiles it has records of. Interpolation is omitted from Alg.~\ref{alg:count} for the sake of brevity. Instead, we assume that there is at least one value associated with each element of the input stream. However, our published implementation is able to correctly prune the sliding window and update its clusters when it detects such a skip in the period counter.

\subsection{Retaining data points with S-RASTER$'$}
\label{sec-prime}
There are use cases where it is desirable to not only identify clusters based on their significant tiles but also on the data points that were projected to those tiles (cf.~Figs.~\ref{fig:idea3} and~\ref{fig:idea4}). In the following, we refer to the variant of S-RASTER that retains relevant input data as S-RASTER$'$. The required changes are minor. With the goal of keeping the overall design of the algorithm unchanged, first the~$\mathbold{\pi}$ nodes have to be modified to produce a stream of tuples $(x_\pi, y_\pi, x, y, \Delta_z)$, i.e.~it retains the original input coordinates. In the $\mathbold{\alpha}$ nodes the hashmaps \emph{totals} and \emph{window} have to be changed to retain multisets of unscaled coordinates $(x, y)$ per projected pair $(x_\pi, y_\pi)$. Counts are given by the size of these multisets. This assumes that determining the size of a set is an $\mathcal{O}(1)$ operation in the implementation language, for instance due to the underlying object maintaining this value as a variable. In case a tile becomes significant, each $\alpha$-node sends not just the tile $(x_\pi, y_\pi)$ but also a bounded stream to $\kappa$ that includes all coordinate pairs $(x, y)$ that were projected to it up to that point in the current window. Furthermore, after a tile has become significant, every additional point $(x_\pi, y_\pi)$ that maps to it also has to be forwarded to $\kappa$, which continues for as long as the number of points surjected to that tile meet the threshold $\tau$. Lastly, in the $\kappa$ node, the set \emph{tiles} has to be turned into a hashmap that maps projected tiles $(x_\pi, y_\pi)$ to their corresponding points $(x, y)$ and the output stream has to be modified to return, for each point $(x, y)$ that is part of a cluster, the tuple $(\Delta_j, \mathit{cluster\_id}, x_\pi, y_\pi, x, y)$.

\section{Theoretical Evaluation}
\label{eval-theory}
As we have shown previously, RASTER generates results more quickly than competing algorithms~\cite{ulm2019contraction}, with the caveat that the resulting clusters might not include elements at the periphery of clusters that other methods might include. Yet, the big benefit of our algorithm is its very fast throughput, being able to process an arbitrary amount of data in linear time and constant memory with a single pass. This makes it possible to use a standard workstation even for big data, while comparable workloads with other algorithms would necessitate a much more expensive and time-consuming cloud computing setup; transferring terabytes of data to a data center alone is not a trivial matter, after all, even if we ignore issues of information sensitivity~\cite{Kaisler2013}~\cite{Mazumdar2019}. The same advantages apply to S-RASTER as it is an adaptation of RASTER that does not change its fundamental properties. Our reasoning below shows that the performance benefits of RASTER for data batches, i.e.~linear runtime and constant memory, carry over to S-RASTER for evolving data streams.

\subsection{Linear Runtime} RASTER is a single-pass linear time algorithm. The same is true for S-RASTER, which we will show by discussing the nodes $\pi$, $\alpha$, and $\kappa$ in turn. In general, data is continually processed as a stream, in which every input is only processed once by nodes $\pi$ and $\alpha$. These nodes perform a constant amount of work per incoming data point. The clustering node $\kappa$, however, performs two different actions. First, is continually updates the multiset of significant tiles $\sigma$ and, second, it periodically clusters $\sigma$. It is easy to see that there is at most one update operation per incoming data point, i.e.~changing the multiplicity of the associated value of an incoming point in $\sigma$. In practice, the number of significant tiles $m$ is normally many orders of magnitude smaller than the number of input data points $n$. Furthermore, the threshold $\tau$ for a significant tile is normally greater than 1. Yet, even in the theoretically worst case it is not possible that $m > n$ as this would mean that at least one point was projected to more than one tile, which is not possible as the projection node $\pi$ performs a surjection. Consequently at worst there are $m = n$ significant tiles if there is exactly one point per significant tile and $\tau =1$.

Clustering in $\kappa$ does not happen continually but periodically, i.e.~in an offline manner. Furthermore, clustering significant tiles is a very fast procedure with a very modest effect on runtime. Even though Alg.~\ref{alg:raster}, lls.~12 -- 24, show a nested loop for clustering, this part is nonetheless linear because the inner loop removes elements from the input the outer loop traverses over. Thus, the entire input, which is the set of significant tiles $\sigma$, is traversed only once. In addition, lookup is a $\mathcal{O}(1)$ operation because the location of all neighboring tiles is known.

In the node $\kappa$, clustering is performed once per period, so take the number of input points per period $n$ and the number of significant tiles $m$. As we have shown, it is not possible that $m > n$, so at worst $m = n$. Clustering only happens once per period, for a total of $\mathcal{P}$ periods in the stream. Periods are not expressed in relation to $n$ but are dependent on time. Thus, $\mathcal{P}$ can be treated as a constant. The total runtime complexity of S-RASTER is $\mathcal{O}(n)$. Nodes $\pi$ and $\alpha$ perform a constant amount of work per data point. The same applies to the updating of multiplicities in $\kappa$, which is likewise a constant amount of work per data point. Lastly, the cost of clustering, which is an $\mathcal{O}(m)$ operation, is amortized over all data points in a period, which is a constant because clustering happens only periodically as opposed to continually, i.e.~for each new data point. Expressed as a function on $n$, the time complexity of S-RASTER is, for $\pi$, $\alpha$, and the two parts of $\kappa$, $\mathcal{O}(n) + \mathcal{O}(n) + (\mathcal{O}(n) + \mathcal{O}(\mathcal{P}m)) = \mathcal{O}(n)$. On a related and more practical note, the performance impact of periodic clustering in the $\kappa$ node can be mitigated by running this node on a separate CPU core, which is straightforward in the context of stream processing. This does not affect the previous complexity analysis, however.

\subsection{Constant Memory}
RASTER needs constant memory $M$ to keep track of all counts per tile of the entire (finite) grid. In contrast, S-RASTER uses a sliding window of a fixed length $c$. In the worst case, the data that was stored in $M$ with RASTER requires $cM$ memory in S-RASTER, which is the case if there is, for each discrete period $\Delta_z$, at least one projected point per tile for each tile. Thus, S-RASTER maintains the key property of RASTER of using constant memory.

\section{Experiment}
\label{sec-exp}

This section describes the design of our experiments, gives details of the experimental environment, and specifies how we performed a comparative empirical evaluation of S-RASTER and related clustering algorithms for data streams.

\subsection{Design}
For our experiments, we generated input files containing a fixed number of points arranged in dense clusters. These data sets contain no noise. The reason behind this decision is that RASTER is not affected by it. Noise relates to tiles that contain less than $\tau$ projected data points, which are simply ignored. In contrast, other algorithms may struggle with noisy data, in particular if they retain the entire input, which would potentially disadvantage them. Our primary goal was to measure clustering performance, i.e.~speed. In addition, we took note of various standard clustering quality metrics such as the silhouette coefficient and distance measurements. Concretely, in the first experiment, the chosen algorithms process 5M data points. This set contains 1000 clusters. Every 500K points, we measure how long that part of the input data took to process. This implies a tumbling window, and for each batch of the input there are 100 different clusters, modeling an evolving data stream. This experiment is run ten times. In contrast, in the second experiment, we use a smaller input data set of 2K points. The variant of our algorithm for this experiment is SW-RASTER, which, in contrast to S-RASTER, does not have a notion of time but instead uses points to define window sizes. This modification was done because the algorithms we use for comparisons define the sliding window similarly. In this experiment, we use a number of established clustering quality metrics, i.e.~the within-cluster sum of squares (SSQ)~\cite[p.~26]{hahsler2017introduction}, adjusted Rand index (cRand)~\cite{hubert1985comparing}, silhouette coefficient~\cite{ROUSSEEUW198753}, and Manhattan distance.

\subsection{Data Set}
S-RASTER was designed for handling a particular proprietary real-world data set, which we cannot share due to legal agreements. However, in order to evaluate this algorithm, we developed a data generator that can create an arbitrarily large synthetic data set that has similar properties. This data generator is available via our code repository. In short, the data generator randomly selects center points on a 2D plane and scatters points around each such center point. There is a minimum distance between each center. For the experiment, we created a file with 1000 clusters of 500 points each, i.e.~500K points in total per batch. A batch corresponds to a period, e.g.~one day. As we have processed ten batches, the total is 5M data points.

\subsection{Environment}
The used hardware was a workstation with an Intel Core i7-7700K and 32 GB RAM. Its main operating system is Microsoft Windows 10 (build 1903). However, the experiments were carried out with a hosted Ubuntu 16.04 LTS operating system that was executed in VirtualBox 5.2.8, which could access 24 GB RAM.

\subsection{Comparative Empirical Evaluation}
\label{eval-practice}
In order to compare S-RASTER to standard algorithms within its domain, we implemented the algorithm for use in Hahsler's popular R package \texttt{stream}~\cite{hahsler2017introduction}. For our purpose, the main benefit is that it contains an extensive suite for the evaluation of algorithms. Of the available algorithms in this R package, we selected DStream~\cite{chen2007density}, DBstream~\cite{bar2014large}~\cite{Hashler7393836}, and Windowed $k$-means Clustering, which was implemented by Hahsler himself. They were chosen because they are standard clustering algorithms for data streams.

DStream is a density-based clustering algorithm that uses an equally spaced grid. It estimates the density of each cell in a grid, which corresponds to a tile in our description of S-RASTER. For each cell, the density is computed based on the number of points per cells. Subsequently, they are classified as dense, transitional, or sporadic cells. A decaying factor ensures that cell densities reduce over time if no new points are encountered. DBStream uses a dissimilarity metric for data points. If a data point in the incoming data stream is below the given threshold value for dissimilarity of any of the hitherto identified micro-clusters, it is added to that cluster. Otherwise, this data point is the seed of a new cluster. Lastly, Windowed $k$-means clustering is an adaptation of the well-known $k$-means algorithm. It partitions the input based on a fixed number $k$ of seeds.

For each algorithm, we chose parameters that delivered good clustering results, based on visual inspection. Concretely, this led to the following parameter values: DStream uses a grid size of 0.0003. DBStream uses a radius $r = 0.0002$, a minimum weight $\mathit{Cm} =0.0$, and a gap time of 25000 points. Windowed $k$-means uses a window length of 100 and $k = 100$. Any parameter we did not specify but is exposed via the \texttt{stream} package was used with its defaults. Lastly, for S-RASTER, we used a precision $\xi = 3.5$ and a window size $c = 10$. 

\section{Results and Discussion}
\label{sec-eval}

In this section we present the results of our evaluation of S-RASTER as well as a discussion of these results.

\begin{figure}[t]
\centering
\begin{subfigure}{.45\textwidth}
  \centering
  \includegraphics[scale=0.36,  trim=0.1cm 0.1cm 0.1cm 0.1cm, clip]{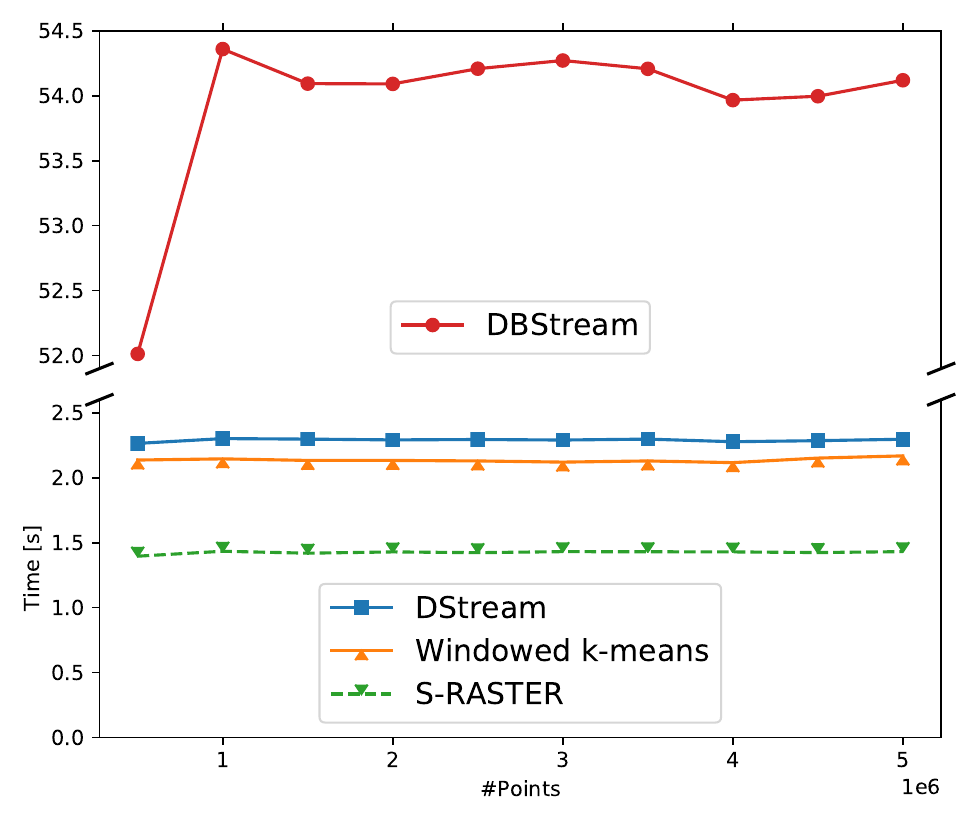}
  \caption{Runtimes}
  \label{fig:time}
\end{subfigure}\hfill
\begin{subfigure}{.45\textwidth}
  \centering
  \includegraphics[scale=0.36, trim=0.1cm 0.1cm 0.1cm 0.1cm, clip]{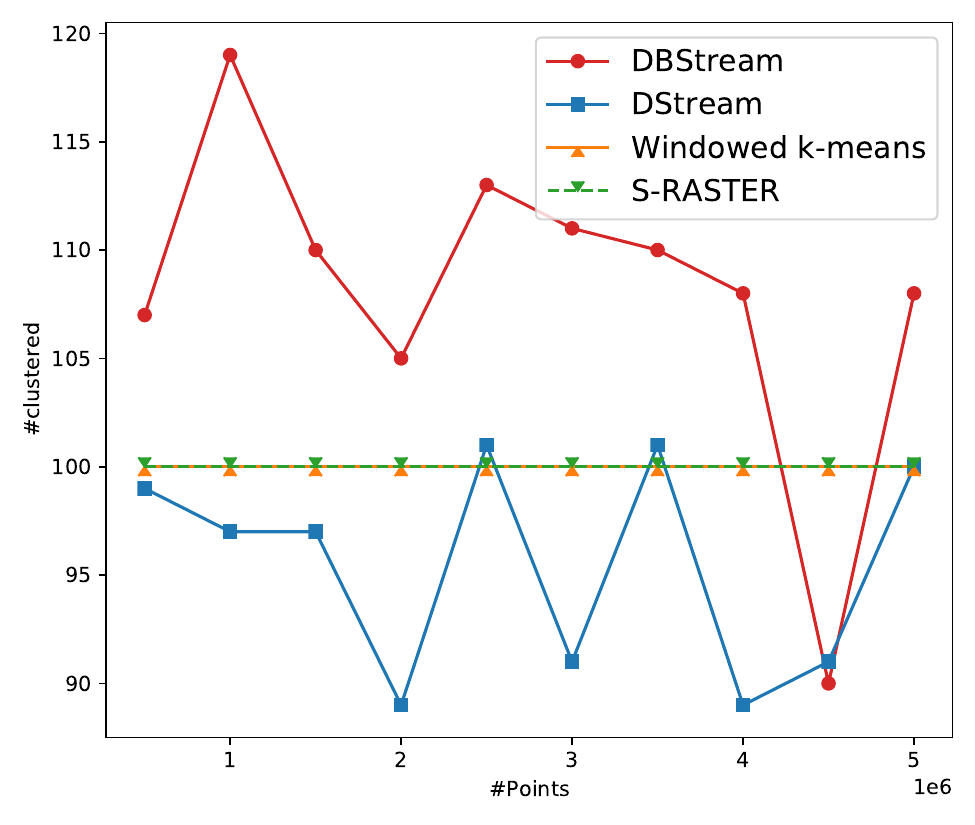}
  \caption{Identified clusters}
  \label{fig:num}
\end{subfigure}\hfill
\caption{Evaluation of S-RASTER (best viewed in color). As Fig.~\ref{fig:time} shows, S-RASTER is faster than competing algorithms in terms of throughput. The algorithms processed 5M points in 10 batches of 500K points. The stated times refer to the end of each batch. As a sanity check, Fig.~\ref{fig:num} plots the number of identified clusters. The input data set contained 100 dense clusters per batch. Windowed $k$-means was provided with the argument $k = 100$. All data is the average of 10 runs.}
\label{fig:throughput}
\end{figure}

\begin{figure}[h]
\centering
\includegraphics[scale=0.60]{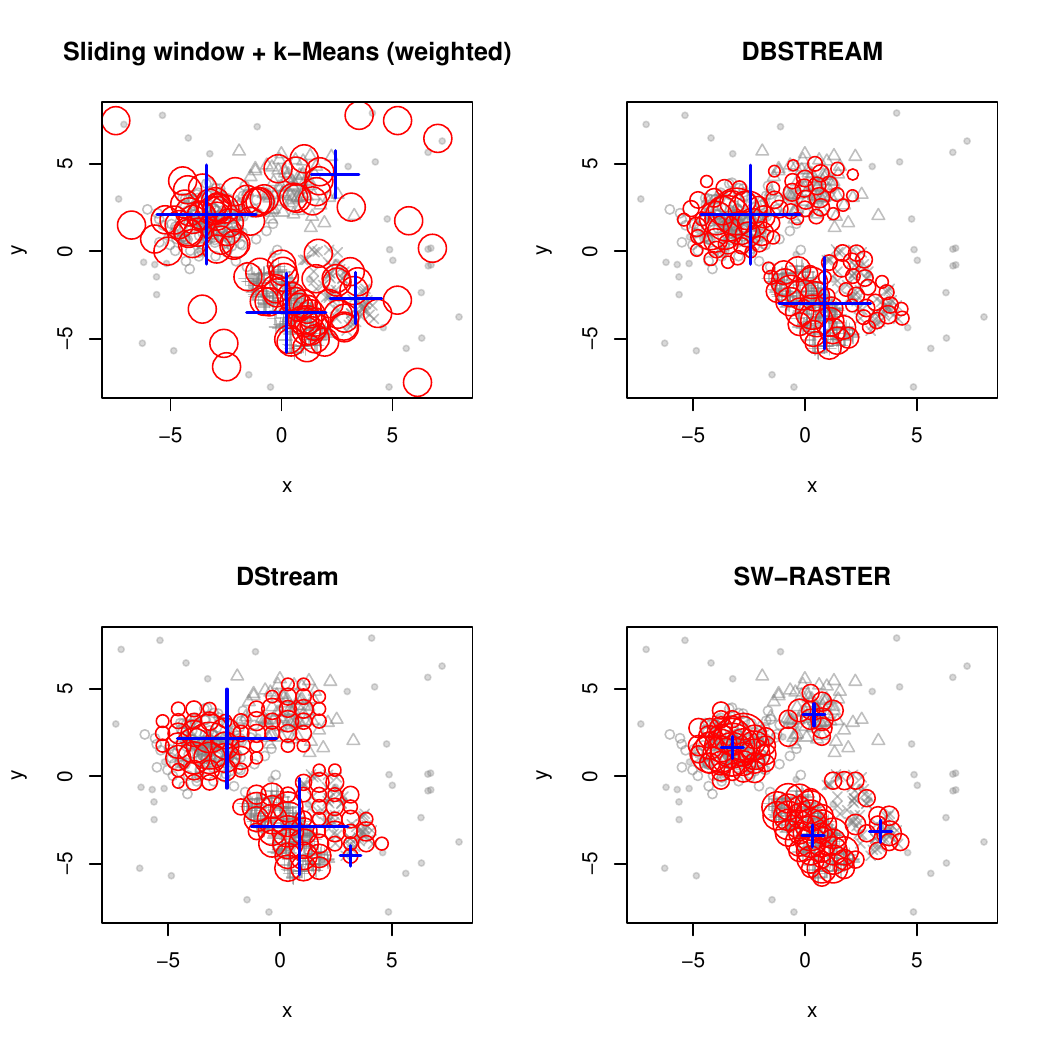}
\caption{Illustration of clustering results with micro and macro clusters after clustering 500 points of the input data stream. S(W)-RASTER produces denser clusters than the other algorithms. The trade-off is that the algorithm ignores points at the periphery of the clusters, depending on the chosen parameters.}
\label{fig:visual}
\end{figure}

\subsection{Results}
The results of the first experiment are shown in Fig.~\ref{fig:throughput}. We start with the most relevant results. As Fig.~\ref{fig:time} shows, S-RASTER is faster than Windowed $k$-means, DBStream and DStream, processing each batch of 500K points in a little less than 1.5s. In contrast, the competing algorithm are at least 50\% slower. DBStream takes about 40 times as long as S-RASTER. Figure~\ref{fig:num} shows the number of clusters the various algorithms have found while processing the input data stream. The Windowed $k$-means algorithm was provided with an argument specifying $k = 100$. S-RASTER and Windowed $k$-means Clustering reliably identify 100 clusters per batch in the input data set whereas DBStream and DStream get close.

The results of the second experiment are summarized in Table~\ref{table:comparison}, and visualized in Fig.~\ref{fig:visual}. SW-RASTER, DBStream, and DStream deliver good results. Conceptually, the algorithms in the \texttt{stream} package identify clusters (macro clusters) that are based on smaller micro clusters, which may be defined differently, based on the chosen algorithm, e.g.~squares in a grid or center points of a circle and their radius. Visual inspection seems to suggest that there are four macro clusters in the data set, which are made up of around 100 micro clusters. SW-RASTER delivers the densest and most separate clusters, which is expressed in the lowest adjusted Rand index (cRand) in this comparison. The cRand measure takes a value between 0 and 1. A value of 1 indicates complete similarity of two partitions, a value of 0 the opposite. SW-RASTER has a cRand value of 0.04, while the other clustering algorithms have cRand values of 0.06. In addition, together with Windowed $k$-means, SW-RASTER has the lowest silhouette coefficient with 0.18. Lastly, SW-RASTER has the lowest Manhattan distance of the chosen algorithms.

\begin{table*}[t]
\caption{Comparison of SW-RASTER with various standard clustering algorithms. The best values in this comparison are listed in bold typeface. The number of micro clusters is listed for the sake of completion but we abstain from making a judgment as the resulting macro clusters are more relevant. Our algorithm does well in this comparison, as evinced by the cRand, silhouette coefficient and Manhattan distance values.} 
\vspace{0.3em}
\centering\ra{1.0}\begin{tabular}{@{}lrccc@{}}
\toprule
 & \phantom{ab} SW-RASTER 
 & \phantom{ab} Windowed k-means
 & \phantom{ab} DStream
 & \phantom{ab} DBStream
 \\
\toprule
Macro clusters & \textbf{4}       & \textbf{4}          & 3         & 2\\
Micro clusters  & 103   & 100      & 108     & 118\\
purity               & 0.93   & 0.94    & 0.96     & \textbf{0.97}\\
SSQ                & 77.80 & 114.26 & 50.70  & \textbf{44.72}\\
cRand             & \textbf{0.04}   & 0.06    & 0.06    & 0.06\\
silhouette        & \textbf{0.18}   & \textbf{0.18}    & 0.21    & 0.27\\
Manhattan      & \textbf{0.11}   & 0.12    & 0.13    & 0.13\\
\bottomrule\end{tabular}
\label{table:comparison}
\end{table*}

\subsection{Discussion}
We have shown that both in theory and practice the benefits of RASTER are retained for S-RASTER. S-RASTER is very fast, outperforming other clustering algorithms for data streams. Of course, the drawback is that there is some loss of precision. In other words, the comparatively good resulting metrics for various cluster quality measures are partly due to the algorithm ignoring points due to the chosen $\delta$ and $\sigma$ parameters. This is also reflected in the lower values for the SSQ and purity metrics, which are due to the algorithm ignoring points at the periphery. That being said, S-RASTER performs very well in the use case it has been designed for, which is not negatively affected by the trade-offs we made in the design of this algorithm. Also note that, at least theoretically, S-RASTER can be easily parallelized (cf.~Sect.~\ref{sec-future}). Yet, as the R package \texttt{stream} is a single-threaded library, this is not an angle we have pursued in this paper. After all, we would not have been able to reap any benefits from creating a multi-threaded implementation in this scenario.

It should be pointed out that both $k$-means clustering and S-RASTER consistently identify 100 clusters per batch in the input data stream. In the case of the former, this is due to the provided parameter $k = 100$, which invariably leads to the identification of 100 clusters. However, the case is much different with S-RASTER. The reason this algorithm identifies 100 clusters in each batch is that this algorithm was developed for reliably detecting dense clusters, ignoring noise and the sparser periphery of a collection of points that competing algorithms may classify as being part of the same cluster. Because there are 100 dense clusters in the input, S-RASTER was able to detect that number with suitable parameter values. In fact, it would have been cause of concern for us had this algorithm not reliably detected all clusters.

\section{Related Work}
\label{sec-related}
Two prominent related algorithms we did not consider in this paper are DUC-STREAM~\cite{gao2005incremental} and DD-Stream~\cite{jia2008grid}, which is due to the absence of a conveniently available open-source implementation. DUC-STREAM performs clustering based on dense unit detection. Its biggest drawback, compared to S-RASTER, is that it has not been designed for handling evolving data streams. While it also uses a grid, the computations performed are more computationally intensive than the ones S-RASTER performs, many of which are based on $\mathcal{O}(1)$ operations on hash tables. DD-Stream likewise performs density-based clustering in grids and likewise uses computationally expensive methods for clustering. DD-Stream is not suited for handling big data. Furthermore, unlike those algorithms, S-RASTER can be effectively parallelized~(cf.~Fig.~\ref{fig:graph}). This primarily refers to the nodes $\pi$ and $\alpha$, which can be executed in an embarrassingly parallel manner.

None of the aforementioned clustering algorithms are quite comparable to S-RASTER, however, as they retain their input. Also, their clustering methods are generally more computationally costly. Thus, S-RASTER requires less time and memory. S-RASTER$'$ is closer to those algorithms as it retains relevant input data. As it does not retain all input, S-RASTER is only inefficient with regards to memory use in highly artificial scenarios. This is the case where there is at most one point projected to any square in the grid, which implies that the algorithm parameters were poorly chosen as the basic assumption is that many points are surjected to each significant tile. Ignoring pathological cases, it can thus be stated that S-RASTER is very memory efficient. Furthermore, clustering, in the $\kappa$ node, is a very fast operation. In summary, S-RASTER is, for the purpose of identifying density-based clusters in evolving data streams, more memory efficient than competing algorithms, and also less computationally intensive. Thus, it is a good choice if the trade-offs they make are acceptable for a given use case.

There is a superficial similarity between the output of self-organizing maps (SOMs)~\cite{kohonen1982self} and S-RASTER. We therefore want to clearly highlight the differences. First, SOMs belong to an entirely different category of algorithms, i.e.~ artificial neural networks (ANNs). They are likewise used for unsupervised learning, albeit there are adaptations for unsupervised online learning~\cite{FURAO2007893}. SOMs were initially developed for visualizing nonlinear relations in high dimensional data~\cite{Kohonen2001}, but they have been applied to clustering problems~\cite{Vesanto2000}~\cite{KIANG2001161} and even suggested as a substitute for $k$-means clustering~\cite{baccao2005self}. Practical clustering applications include, for instance, biomedical analyses~\cite{KOHONEN201352} and water treatment monitoring~\cite{garcia2004}. It may be that SOMs can achieve results similar to S-RASTER, but at an arguably much larger runtime and memory cost, given that distance matrices are the standard data structure and the fact that the so-called Best Matching Unit is determined by computing the minimum Euclidian distance between the input and the neuron weights. Thus, a standard SOM requires $\mathcal{O}(m)$ distance computations for each input point, where $m$ is the number of neurons in the ANN. In contrast, S-RASTER requires only a single projection operation for each input, plus the amortized cost of clustering at the end of each period. Lastly, it is not obvious how sliding windows would be represented with SOMs.

\section{Future Work}
\label{sec-future}
This paper is accompanied by an implementation of S-RASTER for use in the R package \texttt{stream}. In the future, we may release an implementation of S-RASTER for use in Massive Online Analysis (MOA)~\cite{bifet2010moa}. In addition, we may release a complete stand-alone implementation of this algorithm that can be fully integrated into a standard stream processing engine such as Apache Flink~\cite{carbone2015apache} or Apache Spark~\cite{zaharia2010spark}~\cite{zaharia2016apache}. This is particularly relevant for an area we have not considered in this paper, i.e.~the scalability of S-RASTER on many-core systems. As we have shown theoretically, S-RASTER is easily parallelizable. What is missing is to quantify the performance gains that can be expected. The reason we have not pursued this yet is that S-RASTER performs real-world workloads easily in sequential operation. Furthermore, we are interested in applying S-RASTER to data with higher dimensionality. As we elaborated elsewhere~\cite[Sect.~3.3]{ulm2019contraction}, there are $2d$ lookups per dimension $d$. As we are primarily interested in processing 3D data, the additional total overhead due to lookup is modest. As S-RASTER was designed for solving a particular real-world problem, the theoretical objection of the curse of high dimensionality is not relevant.

As S-RASTER was developed in response to a concrete use case (cf.~Sect.~\ref{bg-problem}), we focussed on periodical clustering as this entailed only a negligible cost, in particular because clustering can be performed at convenient times~(cf.~Sect.~\ref{sec-implementation}). However, other use cases may necessitate continual clustering. Thus, it seems worthwhile to explore modifications to the clustering node $\kappa$ that perform the clustering operations in a more efficient manner. We explored a few possibilities for parallelizing of the clustering algorithm in our work on batch processing with RASTER~\cite[Sect.~3.5]{ulm2019contraction}, which would be a good starting point. Another promising idea would be to only selectively perform clustering. Right now, the entire set of significant tiles $\sigma$ is clustered at the end of a period. Yet, in real-world scenarios, it is quite likely that there are not many changes between periods, in particular if they are short. This is even more relevant when we consider the case of continual clustering. In those cases, a lot of redundant work would be performed if we clustered all of $\sigma$, considering that most clusters would not have changed much.

A potential application domain for RASTER and S-RASTER is image clustering, in particular with a focus on clustering multispectral images, which is a well-established area of research~\cite{TRAN20053}. This would necessitate adding another dimension to the algorithm, i.e.~spectral as well as spatial information. There is a potentially wide domain of applications as multivariate images are very common in some domains. Two very prominent examples are magnetic resonance images and remote sensing images. $k$-means clustering has been successfully applied to this problem domain~\cite{MatIsa}~\cite{Ng}. One issue of $k$-means clustering, however, is its susceptibility to get trapped in local optima~\cite{li2015dynamic}, which is not an issue for S-RASTER. It is also the case that $k$-means clustering is slower than S-RASTER. A particularly fruitful field of application for S-RASTER could be image-change detection, which has seen some interest~\cite{zheng2013using}, as some imprecision can be tolerated as long as changes are reliably detected.

\section{Conclusions}
\label{sec-conclusion}
The key takeaway of our original work on RASTER was that by carefully chosen trade-offs, we are able to process geospatial big data on a local workstation. Depending on the use cases, those trade-offs may furthermore have a negligible impact on the precision of the results. In fact, in the case of the problem of identifying hubs in a batch of geospatial data, the loss of precision is immaterial. However, because RASTER is limited to processing batch data, we redesigned this algorithm as S-RASTER, using a sliding window. Thus, S-RASTER can be used to determine clusters within a given interval of the data in real-time. This algorithm is particularly relevant from an engineering perspective as we retain the same compelling benefits of RASTER, i.e.~the ability to process data in-house, which leads to significant savings of time and cost compared to processing data at a remote data center. It also allows us to sidestep problems related to data privacy as business-critical geospatial data can now remain on-site. The trade-offs of S-RASTER compared to other streaming algorithms are also worth pointing out, as we, again, carefully designed its features with an eye to real-world applications. While many clustering algorithms for data streams continually update the clusters they identified, S-RASTER avoids this overhead by doing so only in fixed intervals, which is made possible by the very fast clustering method of RASTER, entailing an insignificant amortized cost. The overall result is that S-RASTER is very fast and delivers good results. Consequently, this algorithm is highly relevant for real-world big data clustering use cases.

\newcommand{\CC}{C\nolinebreak\hspace{-.05em}\raisebox{.4ex}{\tiny\bf +}\nolinebreak\hspace{-.10em}\raisebox{.4ex}{\tiny\bf +}}

\section*{Declarations}
\subsection*{Availability of data and materials}
The datasets generated and analyzed during the current study are available in the following source code repository: \url{https://github.com/FraunhoferChalmersCentre/s-raster}. This repository contains a data generator and the entire code used for benchmarking the algorithm, including a complete implementation of S-RASTER. Additionally, we provide a reference implementation of RASTER in Python as well as a proof-of-concept implementation of S-RASTER in Kotlin.
\subsection*{Competing interests}
The authors declare that they have no competing interests.
\subsection*{Funding}
This research was supported by the project \emph{Fleet telematics big data analytics for vehicle usage modeling and analysis} (FUMA) in the funding program \emph{FFI: Strategic Vehicle Research and Innovation} (DNR 2016-02207), which is administered by VINNOVA, the Swedish Government Agency for Innovation Systems.
\subsection*{Authors' contributions}
Gregor Ulm conceived the S-RASTER algorithm and its initial mathematical formulation, created a prototypal implementation of the algorithm in Kotlin, and wrote the manuscript. Simon Smith ported the existing implementation of S-RASTER to \CC~and conducted experiments. He was assisted by Adrian Nilsson who was also involved in the literature review. All authors were involved in the experimental design and interpretation of results. Mats Jirstrand contributed to the mathematical formulations in this paper. All authors read and approved the final manuscript.
\subsection*{Acknowledgements}
This research project was carried out in the Fraunhofer Cluster of Excellence \emph{Cognitive Internet Technologies}.

\bibliographystyle{splncs04}

\end{document}